\documentclass[%
superscriptaddress,
preprint,
amsmath,amssymb,
]{revtex4-2}
\usepackage{ulem}
\usepackage{xcolor}
\usepackage{multirow}
\usepackage{graphicx}% Include figure files
\usepackage{dcolumn}% Align table columns on decimal point
\usepackage{bm}% bold math
\usepackage{hyperref}% add hypertext capabilities

\begin{document}

%\preprint{APS/123-QED}

\title{Unusual antiferromagnetic order and fluctuations in RbMn$_{6}$Bi$_{5}$}

\author{Chao Mu}
\affiliation{Beijing National Laboratory for Condensed Matter Physics and Institute of Physics, Chinese Academy of Sciences, Beijing 100190, China}
\affiliation{School of Physical Sciences, University of Chinese Academy of Sciences, Beijing 100190, China}

\author{Long Chen}
\affiliation{Beijing National Laboratory for Condensed Matter Physics and Institute of Physics, Chinese Academy of Sciences, Beijing 100190, China}
\affiliation{School of Physical Sciences, University of Chinese Academy of Sciences, Beijing 100190, China}

\author{Jiabin Song}
\affiliation{Beijing National Laboratory for Condensed Matter Physics and Institute of Physics, Chinese Academy of Sciences, Beijing 100190, China}
\affiliation{Collaborative Innovation Center for Nanomaterials Devices, College of Physics, Qingdao University, Qingdao 266071, China}

\author{Wei Wu}
\affiliation{Beijing National Laboratory for Condensed Matter Physics and Institute of Physics, Chinese Academy of Sciences, Beijing 100190, China}

\author{Gang Wang}
\affiliation{Beijing National Laboratory for Condensed Matter Physics and Institute of Physics, Chinese Academy of Sciences, Beijing 100190, China}
\affiliation{School of Physical Sciences, University of Chinese Academy of Sciences, Beijing 100190, China}

\author{Jinguang Cheng}
\affiliation{Beijing National Laboratory for Condensed Matter Physics and Institute of Physics, Chinese Academy of Sciences, Beijing 100190, China}
\affiliation{School of Physical Sciences, University of Chinese Academy of Sciences, Beijing 100190, China}

\author{Zheng Li}
\email{lizheng@iphy.ac.cn}
\affiliation{Beijing National Laboratory for Condensed Matter Physics and Institute of Physics, Chinese Academy of Sciences, Beijing 100190, China}
\affiliation{School of Physical Sciences, University of Chinese Academy of Sciences, Beijing 100190, China}

\author{Jianlin Luo}
\email{jlluo@iphy.ac.cn}
\affiliation{Beijing National Laboratory for Condensed Matter Physics and Institute of Physics, Chinese Academy of Sciences, Beijing 100190, China}
\affiliation{School of Physical Sciences, University of Chinese Academy of Sciences, Beijing 100190, China}

\begin{abstract}
Quasi-one-dimensional RbMn$_{6}$Bi$_{5}$, the first pressure-induced ternary Mn-based superconductor, exhibits a phase diagram analogous to those of cuprate and iron-based superconductors, with superconductivity neighboring antiferromagnetic order. Here, we use $^{55}$Mn and $^{87}$Rb nuclear magnetic resonance (NMR) to unravel its magnetic structure and fluctuations. Above the N\'eel temperature ($T_{\rm N}$), strong antiferromagnetic fluctuations dominate, characteristic of a paramagnetic state with pronounced spin-lattice relaxation rate enhancement. Below $T_{\rm N}$, a first-order phase transition establishes a commensurate antiferromagnetic order, where Mn atoms at the pentagon corners exhibit distinct magnetic moments with different orientations, while the central Mn atom carries no magnetic moment. The complex magnetic architecture, revealed by zero-field and high-magnetic-field NMR spectra, contrasts with earlier neutron diffraction models proposing uniform spin density waves, instead supporting localized moments ordering with charge rearrangement. The proximity of robust antiferromagnetic fluctuations to the high-pressure superconducting phase suggests a potential role for magnetic excitations in mediating unconventional Cooper pairing, akin to paradigmatic high-$T_c$ systems. These findings provide critical insights into the interplay between geometric frustration, magnetic order, and superconductivity in manganese-based materials.
\end{abstract}

%%% Keywords. ?????
%\keywords{NMR, antiferromagnetic, superconductivity}

%\PACS{74.25.Ha, 76.60.-k, 76.60.Gv}

\maketitle

\section{Introduction}
$3d$-transition-metal compounds often show a variety of charge and magnetic orderings due to their unique electronic structure. Particularly, the discovery of high-temperature superconductivity in cuprates and iron-pnictides has spurred researchers to explore unconventional superconductivity in the $3d$-transition-metals\cite{Bednorz1986,Kamihara2008}. Subsequently, the newly discovered nickel-based superconductors also demonstrate the importance of $3d$ electrons in high-temperature superconductivity\cite{Li2019Ni,WangM2023,WangM2024,WenHH2022}. In their phase diagram, superconductivity is adjacent to a magnetic phase and the magnetic fluctuations are believed to be the binding glue of Cooper pairs \cite{Keimer2015,Shibauchi2013}. Therefore, comprehending the magnetic phase adjacent to superconductivity is essential.

Recently, superconductivity was observed in $A$Mn$_{6}$Bi$_{5}$ under high pressure, with $A=$ K, Rb, and Cs \cite{KMB2022SC, Yang2022, long2022} whereas NaMn$_{6}$Bi$_{5}$ does not show bulk superconductivity\cite{Shan2023}. These materials possess a quasi-one-dimensional monoclinic structure featuring infinite [Mn$_6$Bi$_5$]-columns, which are made up of stacked pentagons of Mn atoms\cite{Bao2018,Jung2019,Chen2021Rb,Zhou2022Na}. The pentagon with 5-fold rotational symmetry, which usually exists in quasi-crystals, gives rise to a remarkable frustration effect\cite{Bao2022}, where competing magnetic interactions prevent the system from achieving a simple, ordered magnetic state. This is why antiferromagnetic ordered quasi-crystals are rare\cite{Tamur2025}. At ambient pressure, all $A$Mn$_{6}$Bi$_{5}$ compounds exhibit antiferromagnetic (AFM) order at low temperatures\cite{Bao2018,Jung2019,Chen2021Rb,Zhou2022Na}. However, when high pressure is applied, the magnetic order is suppressed and superconductivity emerges\cite{Zhou2023}. Interestingly, the proximity of superconductivity to a magnetic instability in the phase diagram resembles that in cuprates and iron-pnictides, suggesting an unconventional pairing mechanism mediated by magnetism\cite{KMB2022SC}. Hence, it is urgent to clarify the magnetic correlations of $A$Mn$_{6}$Bi$_{5}$. Moreover, unraveling the pentagonal antiferromagnetic structure could yield profound insights into the compatibility of unconventional symmetries with magnetic order.

In this work, we report nuclear magnetic resonance (NMR) measurements on high-quality RbMn$_{6}$Bi$_{5}$ in both zero-field and high-magnetic-field. Our results reveal that RbMn$_{6}$Bi$_{5}$ exhibits a complex magnetic structure characterized by a first-order phase transition, where the Mn atoms at the corners of the pentagon exhibit different magnetic moments, while the Mn atom at the center of the pentagon does not possess a magnetic moment. Strong antiferromagnetic fluctuations are detected via NMR measurements on the central Mn and Rb.

\section{Materials and method}
Single crystals of $A$Mn$_{6}$Bi$_{5}$, with $A=$ Na, K, Rb, and Cs, were grown by the high-temperature solution growth method\cite{Chen2021Rb}. NMR measurements were performed using a phase coherent spectrometer from Thamway Co. Ltd. The gyromagnetic ratios are $^{55} \gamma = 10.554622$ MHz/T for $^{55}$Mn and $^{87} \gamma = 13.931541$ MHz/T for $^{87}$Rb\cite{Harris2008}. The external magnetic field at the sample position was calibrated by the NMR copper coil. The spectra were obtained by a conventional Hahn spin-echo technique with a typical $\pi/2$ pulse length of $5\ \mu$s for Mn and $3\ \mu$s for Rb. The spin-lattice relaxation time $T_{1}$ was measured at the central peak using a single saturation pulse\cite{Narath1967}. It was determined from a fit of the nuclear magnetization to $1-M(t)/M(\infty) = \frac{1}{35} e ^{-t/T_1}+\frac{8}{45} e ^{-6t/T_1}+\frac{50}{63} e ^{-15t/T_1}$ for Mn and $1-M(t)/M(\infty) = \frac{1}{10} e^{-t/T_1} + \frac{9}{10} e^{-6t/T_1}$ for Rb, where $M(t)$ is the nuclear magnetization at time $t$ after the saturation pulse.

\section{Result and discussions}

\begin{figure*}%[H]
\includegraphics[width=0.8\textwidth,clip]{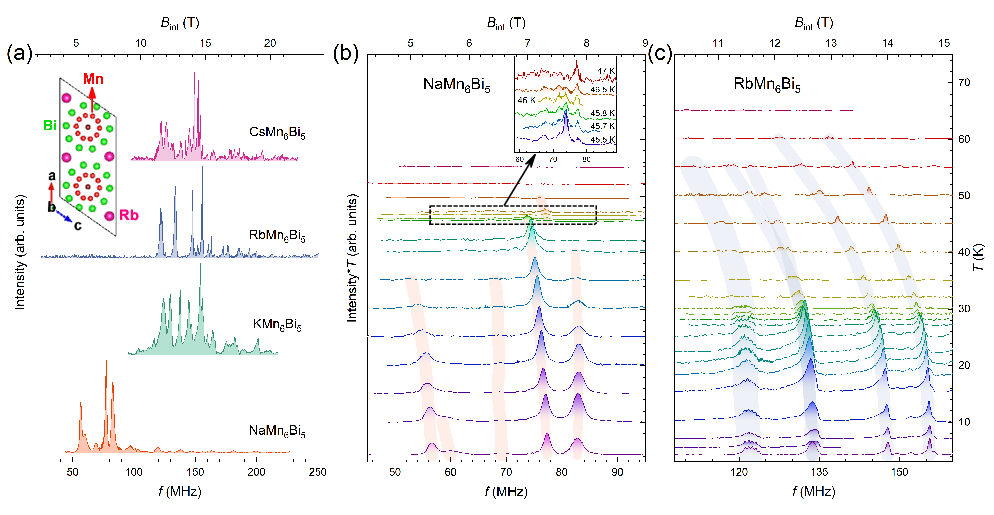}
\caption{Zero field $^{55}$Mn NMR spectra. (a) The spectra of $A$Mn$_{6}$Bi$_{5}$ with $A=$ Cs, Rb, K, and Na at $4.2$ K. The inset illustrates the crystal structure of RbMn$_{6}$Bi$_{5}$. (b) The temperature-dependent spectra of NaMn$_{6}$Bi$_{5}$. The inset is the enlarged view of the coexistence zone. (c) The temperature-dependent spectra of RbMn$_{6}$Bi$_{5}$. The baselines of all curves are shifted to the value of their temperature on the right-hand ordinates. The spectra are scaled by intensity multiplied by temperature. The internal field, proportional to the resonance frequency, is labeled at the top.}\label{fig:f0T}
\end{figure*}

Figure \ref{fig:f0T}(a) shows the zero-field NMR spectra of $A$Mn$_{6}$Bi$_{5}$ at $4.2$ K with $A=$ Cs, Rb, K, and Na, respectively. The resonance frequencies are the product of the gyromagnetic ratio $^{55}\gamma$ and the internal magnetic field $B_{\rm int}$, as given by $f =\ ^{55}\gamma B_{\rm int}$. $B_{\rm int}$ can be directly read out from the peak position, which is indicated by the top label. All samples display discrete resonance peaks, indicating a commensurate magnetic order. In addition to the main resonance peaks associated with the magnetic order, small peaks at high frequencies should originate from manganese compound impurities\cite{Shimizu2006,Manago2022}, which are much smaller than the main peaks and only show up at low temperatures. The internal fields are similar for CsMn$_{6}$Bi$_{5}$, RbMn$_{6}$Bi$_{5}$, and KMn$_{6}$Bi$_{5}$ samples, while the internal magnetic field of NaMn$_{6}$Bi$_{5}$ is half that of them. The spectra of NaMn$_{6}$Bi$_{5}$, shown in Fig. \ref{fig:f0T} (b), change when the temperature crosses $46$ K, indicating a first-order magnetic transition characterized by the coexistence of two phases. Both heat capacity measurements and magnetic susceptibility measurements have shown two transitions at $47.3$ K and $51.8$ K\cite{Zhou2022Na}. The characteristics of NaMn$_{6}$Bi$_{5}$ are complex and pose significant challenges to current research efforts. Given these challenges, we will leave this interesting issue for future study.

RbMn$_{6}$Bi$_{5}$, exhibiting simpler spectra compared to KMn$_{6}$Bi$_{5}$ and CsMn$_{6}$Bi$_{5}$, was chosen for further analysis, as shown in Fig. \ref{fig:f0T} (c). Below $T_{\rm N}$, five distinct humps emerge as the temperature decreases, with the second and the third humps merging at lower temperatures. Each hump contains five peaks attributed to quadrupole splitting, which will be discussed later. The spectra shape may be asymmetric when these five peaks overlap\cite{Manago2022}. The presence of five humps indicates five non-equivalent Mn sites with varying internal magnetic fields, suggesting that the magnetic order is commensurate. If the magnetic order is incommensurate, the spectrum should be much broader, lacking distinct peaks\cite{Vinograd2021}. At $4.2$ K, these fields, calculated as $B_{\rm int}=f/^{55}\gamma$, are $11.5$ T, $12.6$ T, $12.7$ T, $14$ T, and $14.7$ T. These values can be read from the top label of Fig. \ref{fig:f0T}(c). The magnitude of the largest field is a quarter larger than that of the smallest one. The large difference cannot be due to the hyperfine coupling constant. Given that the internal field is primarily influenced by its electrons, the value of the hyperfine coupling constant should be similar to that of other manganese materials. Assuming the hyperfine coupling constant $7$ T/$\mu_{\rm B}$, which is the value of AFM $\alpha$-Mn\cite{Manago2022}, the magnetic moments at five sites are estimated to be $1.6$ $\mu_{\rm B}$, $1.8$ $\mu_{\rm B}$, $1.8$ $\mu_{\rm B}$, $2.0$ $\mu_{\rm B}$, $2.1$ $\mu_{\rm B}$ respectively. These values are similar to the neutron results of $2.64$ $\mu_{\rm B}$\cite{Bao2022}. However, neutron results demonstrated incommensurate spin density waves (SDW), which conflict with our data because they would broaden the NMR spectra rather than separate the peaks.

\begin{figure*}%[H]
\includegraphics[width=0.8\textwidth,clip]{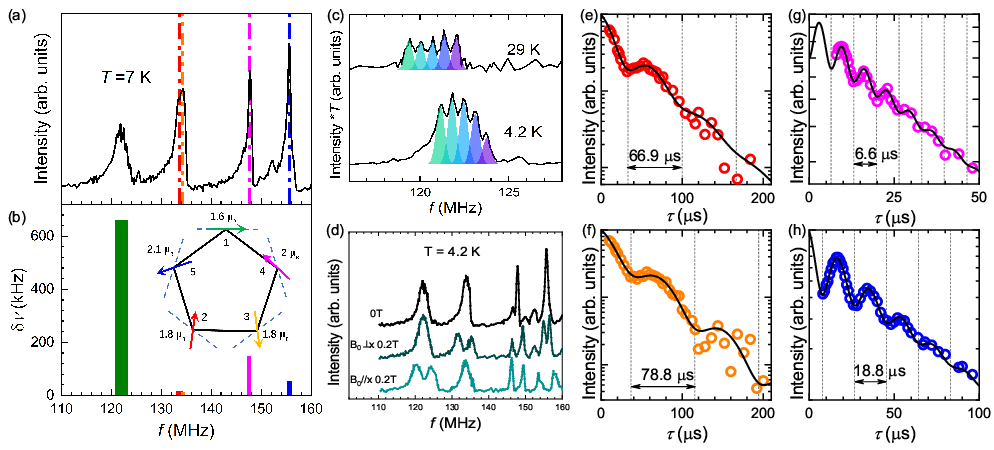}
\caption{\label{fig:nuQ} Magnetic structure is deduced from spectra and quadrupole splitting. (a) The spectrum of $^{55}$Mn at $7$ K. (b) The quadrupole splitting. The line widths are the full width at half maximum of each peak. The inset shows the magnetic moment directions on the pentagon. The dashed lines indicate the calculated EFG principal axes. The labels indicate the peaks from low to high frequency, and the colors represent the relationships between the peaks and the magnetic moments. (c) The enlarged view of the peaks from site 1. There are five peaks due to quadrupole splitting. (d) Comparison of spectra at zero field and an applied $0.215$ T field. (e,f,g,h) The $^{55}$Mn echo amplitude as a function of inter-pulse delay $\tau$ is measured at the frequencies marked by the corresponding colored dash-line in (a). The inverse of oscillation periods give the splitting frequencies which are shown in (b).}
\end{figure*}

$^{55}$Mn with $I=5/2$ should have five resonance peaks when there is an electric field gradient (EFG), like the spectra at around $122$ MHz, as shown in Fig \ref{fig:nuQ} (a) and (c). However, other humps cannot be decomposed into five peaks directly due to small splitting and peak overlap. The splitting frequencies are not only proportional to EFG strength but also affected by the angle between the EFG principal axis and the internal magnetic field. When the magnetic Zeeman interaction dominates over the quadrupole interaction\cite{Kitagawa2008},
\begin{equation} \label{eq:vQ}
\nu_{m \leftrightarrow m-1}  = \gamma B_{\rm int}
 +\frac{\nu_Q}{2}(m-\frac{1}{2})(3 \mathrm{cos} ^2 \theta -1+\eta \mathrm{sin} ^2 \theta \mathrm{cos} 2\phi)
\end{equation}
where $\nu_Q \equiv 3e^2qQ/2I(2I-1)h$ is the quadrupole frequency along the principal axis, $\eta = (V_{xx}-V_{yy})/V_{zz}$ is the asymmetry parameter of the EFG tensor $V_{\alpha\alpha}$, $eq=V_{zz}$, $Q$ is quadrupole and $h$ is Planck constant. $\theta$ and $\phi$ specify the polar angle of the internal magnetic field $B_{\rm int}$ concerning the principal axis\cite{Narita1966}.
The quadrupole interaction results in a splitting of the resonance peak, as shown in Fig. \ref{fig:nuQ} (c). If $\eta = 0$, the quadrupole splitting will vanish at the so-called magic angle $\theta=\arccos(1/\sqrt{3}) \approx 54.7 ^{\circ}$, which is almost half of the pentagon angle $108 ^{\circ}$. Despite the non-negligible impact of $\eta$ due to the low symmetry of space group C2/m, the splitting can also vanish at a specific angle when $3 \mathrm{cos} ^2 \theta -1 + \eta \mathrm{sin} ^2 \theta \mathrm{cos} 2\phi=0$. With $\phi = 0$ since the magnetic moments lie in the pentagon plane, the quadrupole splitting will vanish in the range $54.7 ^{\circ} < \theta < 90 ^{\circ}$. The largest splitting of site 1 indicates that its moment is along the principal axis of the EFG, perpendicular to the radial direction, as shown by the dashed lines in the inset of Fig \ref{fig:nuQ} (b). To extract the tiny splitting of other sites, we performed echo decay measurements on the other four peaks marked by color dash lines in Fig \ref{fig:nuQ} (a), a technique commonly employed to determine spin-spin relaxation times\cite{Jeglic2010}. As shown in Fig. \ref{fig:nuQ}(e - h), the recovery curves oscillate with the periods of $t_{\rm Q}=66.9$, $78.8$, $6.6$, and $18.8 \ \mu$s, respectively. The quadrupole splitting can be calculated as $\delta\nu=1/t_{\rm Q}$ with the corresponding results presented in Fig. \ref{fig:nuQ} (b). We should note that the overlapping peaks of site 2 and site 3 may introduce interference effects in the oscillations, potentially masking the quadrupole splitting. Since the deduced $\delta\nu$ of site 2 and site 3 is not reliable, additional information is needed to determine their angles.

We applied a small magnetic field, $B_{0}=0.215$ T, in two orthogonal directions, $B_0 \parallel x$ and $B_0 \perp x$, with the $x$ direction aligned with the moment of site 1, depicted in Fig \ref{fig:nuQ} (d). When the applied field $B_{0}$ is parallel to the internal field $B_{\rm int}$, the effective magnetic field is $B_{\rm eff}=B_{\rm int}\pm B_{0}$, causing the spectra to split. On the other hand, when $B_{0}$ is perpendicular to $B_{\rm int}$, the effective magnetic field is $B_{\rm eff}=\sqrt{B_{\rm int}^2+B_{0}^2}$, leading to a shift in the spectra towards higher frequencies. Here $B_{\rm int} \sim 12$ T is much larger than $B_{0}=0.215$ T, so the shift is about $0.02$ MHz, which is too small to be discernible. Site 2 and site 3 exhibit the opposite angle-dependent splitting to site 1, indicating their moments are nearly perpendicular to the moment of site 1, and the angle with respect to the $x$-axis is estimated to be $\pm 82^{\circ}$. Site 4 and 5 split in both directions and the angles with respect to the $x$ direction are estimated to be $\pm 42^{\circ}$ and $\pm 19^{\circ}$ respectively. Quadrupole splitting results have indicated that for sites 2, 3, 4, and 5, the angles between the magnetic moment and the EFG are near the magic angle. So, by combining the quadrupole results with the AFM correlation (which will be discussed later), we assign the magnetic moments to the pentagon corner as illustrated in the inset of Fig. \ref{fig:nuQ} (b). It is slightly different from neutron results\cite{Bao2022}, which assume uniform magnetic moment amplitudes for all five Mn atoms. Our analysis reveals distinct magnetic moment amplitudes for each site, with the largest being a quarter greater than the smallest. In addition, there is no resonance peak at low frequency, as shown in Fig. \ref{fig:f0T} (a), indicating the amplitudes of the magnetic moments are not modulated. The magnetic moment directions may rotate in tandem with pentagons or flipped to the opposite direction layer by layer, but the magnetic moment amplitudes remain unchanged.

\begin{figure*}%[H]
\includegraphics[width=0.8\textwidth,clip]{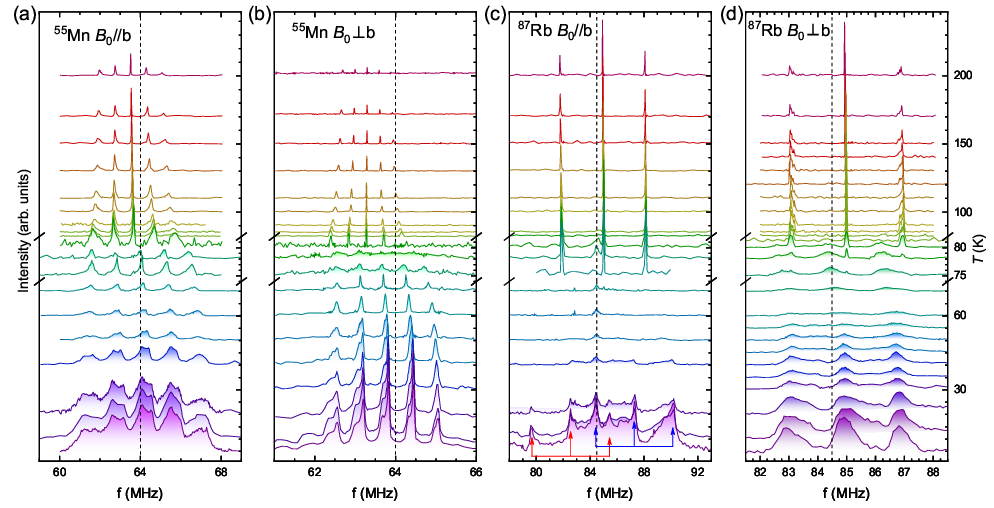}
\caption{\label{fig:f6T} Spectra of the central $^{55}$Mn (a)$B_{0} \parallel b$ and (b)$B_{0} \perp b$. Spectra of $^{87}$Rb (c)$B_{0} \parallel b$ and (d)$B_{0} \perp b$. The baselines of all curves are shifted to the value of their temperature on the right-hand ordinates. The spectral intensity of $^{55}$Mn at $4.2$ K is multiplied by $0.42$ to avoid overlap. The vertical dash-lines mark the frequency of $^{55}\gamma B_{0}$ and $^{87}\gamma B_{0}$, corresponding to $^{55}$Mn and $^{87}$Rb respectively. The spectra of $^{87}$Rb in both directions show the coexistence of the paramagnetic phase and ordered phase in the temperature range from $75$ K to $80$ K, which is shown on an enlarged scale. Below $T_{\rm N}$, the spectra of $^{55}$Mn split for both directions. For $^{87}$Rb, the spectra split only for $B_{0} \parallel b$, with two sets of splitting peaks are marked by red and blue arrows, respectively.}
\end{figure*}

There is another Mn atom situated at the pentagon center and neutron results imply that it has a very small or no magnetic moment\cite{Bao2018}. A magnetic field needs to be applied to measure non-magnetic atoms. Figure \ref{fig:f6T} (a) and (b) show the NMR spectra of $^{55}$Mn at $6.065$ T with $B_{0} \parallel b$ and $B_{0} \perp b$, respectively. The quadrupole splitting parallel to the $b$-axis is approximately double that perpendicular to the $b$-axis, indicating that the principal direction of EFG is the $b$-axis and the spectra originate from the central Mn atom by symmetry. Corner Mn cannot be detected because of its strong magnetic fluctuations and short relaxation times. Below $T_{\rm N}$, the slight shift in the spectra indicates that, if present, the magnetic moment is less than $0.01 \ \mu_{\rm B}$. Consequently, there are two kinds of Mn atoms, one with the magnetic moment at the corner of the pentagon and one with no magnetic moment at the center.

We also measured the NMR spectra of $^{87}$Rb with $I=3/2$, which is located outside [Mn$_6$Bi$_5$]-columns, as shown in Fig. \ref{fig:f6T} (c) and (d). Below $T_{\rm N}$ the spectra split for $B_{0} \parallel b$ and broaden for $B_{0} \perp b$. This indicates that the field at the Rb site is mainly along the $b$-axis and its magnitude is approximately $\pm 0.35$ T. The small internal field at Rb implies that the magnetic field generated by the [Mn$_6$Bi$_5$]-columns is mostly cancelled out at the Rb site and the [Mn$_6$Bi$_5$]-columns are antiferromagnetically correlated with each other.

\begin{figure*}%[H]
\includegraphics[width=0.8\textwidth,clip]{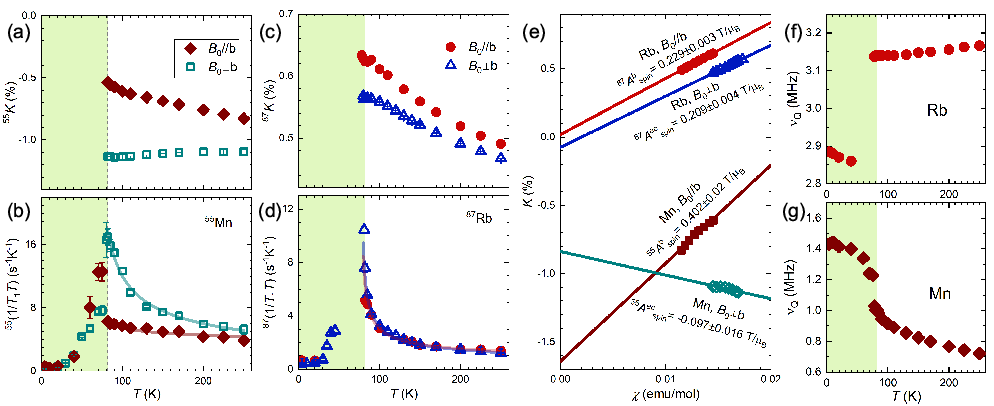}
\caption{\label{fig:T1K} NMR shift and spin-lattice relaxation rate of (a)(b) $^{55}$Mn and (c)(d)  $^{87}$Rb are presented, respectively. The regions corresponding to magnetic order states are marked in green. (e) $K$ is plotted against the bulk susceptibility for both $^{55}$Mn and $^{87}$Rb. The lines represent the fits to a linear relation, and the slopes are the diagonal hyperfine coupling constants. The temperature dependence of $\nu_{\rm Q}$ for (f) Rb and (g) Mn is presented, respectively.}
\end{figure*}

At the central peak of the spectra of $^{55}$Mn and $^{87}$Rb, the NMR shift $K$ and spin-lattice relaxation time $T_{1}$ are measured. The second-order quadrupolar effect is subtracted for $H \perp b$\cite{Huang2022, Lu2023}. $K$ consists of a $T$-independent chemical shift $K_{\rm orb}$ from orbital susceptibility and a $T$-dependent Knight shift $K_{\rm spin}$ due to the uniform magnetic spin susceptibility $\chi(q=0)$. The orbital component $K_{\rm orb}$ is estimated by the intercept of the $K-\chi$ plot, as shown in Fig. \ref{fig:T1K} (e). $K_{\rm orb}$ of Mn is negative for both directions and $K_{\rm orb}$ of Rb is small. By subtracting $K_{\rm orb}$, $K_{\rm spin} = A_{\rm spin} \chi_{\rm spin}$ can be obtained, where $A_{\rm spin}$ is the hyperfine coupling constant. This hyperfine coupling constant is different from that of the Mn at the corner of the pentagon, because the positions of the corresponding atoms are different. We note that $A_{\rm spin}$ of Mn for $H \perp b$ is negative. The decrease of $K_{\rm spin}$ of Mn for $H \perp b$ with decreasing temperature, as shown in Fig. \ref{fig:T1K} (a), means $\chi_{\rm spin}$ increases. So, all $\chi_{\rm spin}$ of Mn and Rb for both directions increase with decreasing temperature.

The $1/T_{1}T$ of Mn and Rb increases with decreasing temperature for both directions, as shown in Fig. \ref{fig:T1K} (b) and (d), which is indicative of strong magnetic fluctuations. Usually, when $\chi_{\rm spin}$ also increases with decreasing temperature, the fluctuations are ferromagnetic (FM) fluctuations\cite{Moriya1963}. However, the increase of $\chi_{\rm spin}$ is small due to frustration and imperfect internal field cancellation. Since the ordered state is AFM rather than FM, the main fluctuations are AFM fluctuations. The $1/T_{1}T$ of both $^{55}$Mn and $^{87}$Rb can be fitted by $1/T_{1}T=C/\sqrt{(T-\theta)}+(1/T_{1}T)_{0}$ for AFM metals\cite{Moriya1985}, as the curves shown in \ref{fig:T1K} (b) and \ref{fig:T1K} (d), where $(1/T_{1}T)_{0}$ is a constant due to noninteracting electrons. Below $T_{\rm N}$, magnetic moments are ordered and the fluctuations are suppressed, which is manifested as a decrease of $1/T_{1}T$, as shown in Fig. \ref{fig:T1K} (b) and (d). A jump just below $T_{\rm N}$ in Fig. \ref{fig:T1K} (b) is due to a first-order transition. This first-order phase transition is also supported by the coexistence of two phases around $T_N$, as shown in Fig. \ref{fig:f6T} (c) and (d), where AFM spectra emerge at new frequencies, rather than shifting.

To gain further insight into the transition, we show the quadrupole frequency of Rb and central Mn deduced from the satellite peaks, as shown in Fig. \ref{fig:T1K} (f) and \ref{fig:T1K} (g). These frequencies are proportional to the EFG. Above $T_{\rm N}$, the EFG of Mn increases, indicating a contraction of Mn tubes. The EFG of Rb decreases gradually with decreasing temperature, indicating a slight increase in the tube spacing. Below $T_{\rm N}$, a rapid increase in quadrupole frequency of Mn signifies a substantial enhancement in EFG, while a decrease in quadrupole frequency of Rb indicates a reduction of EFG. It reflects the rearrangement of charge and is consistent with the magnetic-induced charge order\cite{Bao2022}. Such changes cause the AFM transition to exhibit the first-order behavior.

\section{Conclusions}

By conducting $^{55}$Mn and $^{87}$Rb NMR measurements, we investigate the magnetic behavior of the quasi-one-dimensional ternary Mn-based superconductor RbMn$_{6}$Bi$_{5}$. Above $T_{\rm N}$, strong AFM fluctuations dominate, evidenced by enhanced $1/T_{1}T$. Below $T_{\rm N}$, a first-order phase transition establishes a commensurate AFM order with two distinct Mn sublattices: five pentagon-corner Mn atoms exhibit position-dependent magnetic moments ($1.6 \sim 2.1 \mu _{\rm B}$) with different orientations, while the central Mn atom remains non-magnetic. This structural heterogeneity, revealed by zero-field and high-field NMR spectra, contrasts with prior neutron diffraction models of uniform spin density waves, instead supporting localized moments with charge ordering. Since high pressure can inhibit AFM order and simultaneously induce superconductivity, it is reasonable to ask whether AFM fluctuations contribute to the induction of superconductivity. Answering this question could potentially provide deeper insights into the underlying mechanisms of superconductivity in this class of materials.

\begin{acknowledgments}
The authors would like to acknowledge Prof. Jie Ma of the Shanghai Jiao Tong University for valuable discussions. This work was supported by the National Key Research and Development Program of China (Grant No. 2022YFA1602800, 2022YFA1403903, 2023YFA1607400), the National Natural Science Foundation of China (Grant No. 12134018, 52325201),  and the Strategic Priority Research Program and Key Research Program of Frontier Sciences of the Chinese Academy of Sciences (Grant No. XDB33010100), and the Synergetic Extreme Condition User Facility (SECUF).
\end{acknowledgments}

\end{document}